\documentclass[12pt,a4paper]{article}
\usepackage{jcappub}

\pdfoutput=1 

\usepackage{amsmath,amssymb,cases}

\def\be{{\boldsymbol e}}
\def\bk{{\boldsymbol k}}
\def\bx{{\boldsymbol x}}

\def\cA{{\cal A}}
\def\cL{{\cal L}}
\def\cP{{\cal P}}

\def\ri{{\rm i}}

\def\rf{{\rm f}}
\def\rP{{\rm P}}

\title{Viable inflationary magnetogenesis with helical coupling}

\author{Yuri Shtanov}

\affiliation{Bogolyubov Institute for Theoretical Physics, \\ Metrologichna St.\@ 14-b, Kiev 03143, Ukraine} %
\affiliation{Astronomical Observatory, Taras Shevchenko National University of Kiev, \\ Observatorna St.\@ 3, Kiev 04053, Ukraine} %

\emailAdd{shtanov@bitp.kiev.ua}

\abstract{We consider helical coupling to electromagnetism and present a simple scenario of evolution of the coupling function leading to a viable inflationary magnetogenesis without the problem of back-reaction.  In this scenario, helical magnetic fields of strength of order up to $10^{- 7}\,\text{G}$, when extrapolated to the current epoch, can be generated in a narrow spectral band centered at any reasonable wavenumber by adjusting the model parameters. We discuss implications of this model for baryogenesis, which impose additional constraints on the strength and correlation length of magnetic field.}


\arxivnumber{1902.05894}

\begin{document} 
\maketitle
\flushbottom


\section{Introduction}

Magnetic fields permeate our universe on various spatial scales \cite{KF}.  Spiral galaxies similar to Milky Way host regular magnetic fields of the order of $\mu$G, while distant galaxies exhibit fields of the order of $100~\mu$G \cite{Bernet:2008qp, Wolfe:2008nk}.  On cosmological spatial scales, there exist upper bounds on magnetic fields ranging from $10^{-9}$ to $10^{-7}$ G (all values are comoving), based on the analysis of the cosmic microwave background anisotropy, Big-Bang nucleosynthesis, and small-scale structure formation \cite{Kahniashvili:2010wm}. What is important for the present paper is that there is a strong evidence for the presence of magnetic fields in intergalactic medium, including voids \cite{Tavecchio:2010mk, Ando:2010rb, Neronov:1900zz, Dolag:2010, Taylor:2011}, with lower bounds $B \gtrsim 10^{-16}\,\text{G} \times \text{max}\, \big\{ 1, \sqrt{ \text{Mpc}/\lambda} \big\}$, where $\lambda$ is the correlation length of the field.   Interestingly, in \cite{Pandey:2019glp}, it is argued that a primordial magnetic field of the order $10^{-10}$ G can significantly lower the angular momentum barrier to the formation of direct-collapse black holes. All this makes plausible an idea of cosmological origin of magnetic fields, which are subsequently amplified in galaxies, probably by the dynamo mechanism (see reviews \cite{Grasso:2000wj, Widrow:2002ud, Giovannini:2003yn, Kandus:2010nw, Durrer:2013pga, Subramanian:2015lua}).

One of the attractive possibilities of generating magnetic fields is by doing this on the inflationary stage\,---\,this naturally solves the problem of their coherence length, which can be comparable to the size of the large-scale structure.  To amplify the vacuum electromagnetic field, one needs to violate the conformal invariance of the field equations.  As a simple option \cite{Turner:1987bw, Ratra:1991bn}, one considers a modified gauge-invariant action for the electromagnetic field of the form
\begin{equation}\label{model}
	\cL_{\rm em} = - \frac14 I^2  F_{\mu\nu} F^{\mu\nu} - \frac14 f  F_{\mu\nu} \tilde F^{\mu\nu} \, ,
\end{equation}
where 
\begin{equation}
	\tilde F^{\mu\nu} = \frac12 \epsilon^{\mu\nu}{}_{\alpha\beta} F^{\alpha\beta}
\end{equation}
is the Hodge dual of $F^{\mu\nu}$, and $I$ and $f$ are non-trivial functions of time on the stage of inflation due to their dependence on the background fields such as the inflaton, dilaton \cite{Bamba:2003av} or the metric curvature.  The first term in (\ref{model}) is the so-called kinetic coupling; the function $I$ can be interpreted as describing a time-dependent gauge coupling. The second, parity-violating term, is the helical coupling.  Numerous versions of this model have been under consideration in the literature (see \cite{Durrer:2013pga, Subramanian:2015lua} for recent reviews). 

Scenarios based on the kinetic coupling to electromagnetism meet with the issues of back-reaction and strong gauge coupling \cite{Demozzi:2009fu, Urban:2011bu, BazrafshanMoghaddam:2017zgx}.  Essentially, if one assumes the function $I$ to be monotonically decreasing with time, then it is electric field that is predominantly amplified, causing the problem of back-reaction on inflation and making it difficult to generate magnetic fields of plausible values.  If the function $I$ is monotonically increasing, then magnetic field is amplified predominantly, but here one has to deal with strong coupling since the effective gauge coupling evolves as $e_{\rm eff} \propto I^{-1}$ and is quite large at the early stages of evolution \cite{Demozzi:2009fu}.  Whether this strong-coupling problem is serious remains debatable since all computations done in \cite{Ratra:1991bn} and in subsequent papers are semi-classical, not fully quantum-mechanical, and a fully quantum-mechanical formalism does not yet exist.  

One might try to get around these problems altogether by assuming non-monotonic behavior of the coupling $I$, which initially increases from unity to large values and eventually returns back to unity  \cite{Ferreira:2013sqa, Sharma:2017eps, Sharma:2018kgs}.  However, the requirement of a successful magnetogenesis in such  scenarios places rather stringent constraints on inflation, in particular, demanding unusually low reheating temperatures.  The general difficulty with scenarios of this kind is the evolution law of magnetic field $B \propto I / a^2$ (with $a$ being the scale factor) on the super-Hubble spatial scales.  As $I$ decreases from large values back to unity, the gain in the magnetic field obtained during its preceding growth is lost \cite{BazrafshanMoghaddam:2017zgx}.

Several authors considered the model where couplings of both kind are present in (\ref{model}), evolving in a coherent manner \cite{Caprini:2014mja, Sharma:2018kgs} or by different laws \cite{Ng:2015ewp}.  Such scenarios also typically require sufficiently low scale of inflation for their realization while producing magnetic fields of moderate magnitudes (e.g., of the order of $10^{-16}\, \text{G}$ on Mpc scale for the inflationary energy scale of the order of $100\, \text{TeV}$ in the model of \cite{Caprini:2014mja}).

In this paper, we consider model (\ref{model}) with the standard kinetic term ($I \equiv 1$)  and with a non-trivial helical-coupling function $f$.  Contrary to the case of kinetic coupling, the absolute value of $f$ is of no big significance (since the second term in (\ref{model}) with constant $f$ is topological), which greatly broadens the scope of its possible evolution\,---\,the strong-coupling issue does not arise in this model.  Several typical laws of evolution of $f$ were studied in the literature.   Evolution in the form $f \propto (\log a)^p$ during inflation (arising in the case of linear dependence $f (\varphi) \propto \varphi$ on the inflaton $\varphi$) generically leads to maximally helical magnetic fields with blue power spectrum \cite{Durrer:2010mq, Jain:2012jy, Fujita:2015iga}.  Then the constraints stemming from the considerations of back-reaction and the observational bounds on inter-galactic magnetic fields allow for too little power on the comoving scales of galaxies, clusters and voids to account for the magnetic fields in these objects \cite{Durrer:2013pga}.  Evolution in the form of power law $f \propto a^p$ with $p > 0$ \cite{Campanelli:2008kh} also results in negligible amplification of magnetic field (we show this in the next section).

Scenarios with $f \propto a^p$, $p < 0$, appear to have escaped close attention.  In this paper, we partially fill this gap.  We present a simple model of this sort, with $f \propto a^{-1}$ during some time interval and constant outside this interval, allowing for successful magnetogenesis without the problem of back-reaction.  By adjusting the two free parameters of this model (the duration of the evolution period and the change $\Delta f$ during this period), helical magnetic fields of strength of order up to $10^{- 7}\,\text{G}$ at the current epoch can be generated in a narrow spectral band that can be centered at any reasonable wavenumber.

Helical (hyper)magnetic fields source the baryon number in the hot universe \cite{Giovannini:1997gp, Giovannini:1997eg}, opening up an interesting possibility of baryogenesis \cite{Bamba:2006km, Anber:2015yca, Fujita:2016igl, Kamada:2016eeb, Kamada:2016cnb, Jimenez:2017cdr}.  The requirement of not exceeding the observed baryon number density imposes constraints on the strength and correlation length of maximally helical magnetic field.  We briefly discuss them in this paper.

The paper is organized as follows.  In the next section, we revisit the model in which the helical coupling grows as a positive power of the scale factor and show its failure in producing primordial magnetic fields of appreciable values.  In Sec.~\ref{sec:viable}, we present our model of magnetogenesis and derive the magnitude and spectrum of the generated  electromagnetic fields as functions of its free parameters.  In Sec.~\ref{sec:inflation}, we estimate the effect of back-reaction of the electromagnetic energy-density on inflation. In Sec.~\ref{sec:inflaton}, we provide a specific realization of our model for the simplest inflation based on a massive scalar field and study the effect of back-reaction of the electromagnetic field on the inflaton dynamics.  In Sec.~\ref{sec:baryogenesis}, we discuss the implications of our model for baryogenesis. In Sec.~\ref{sec:summary}, we summarize or results.

\section{Power-law growth of the coupling}

In this paper, we consider the model with the Lagrangian\footnote{Throughout the paper, we are using units in which $c = \hbar = 1$.}
\begin{equation}\label{Lem}
	\cL_{\rm em} = - \frac14 F_{\mu\nu} F^{\mu\nu}  - \frac14 f F_{\mu\nu} \tilde F^{\mu\nu} \, ,
\end{equation}
for electromagnetic field, in which $f$ is a function of time through its dependence on the background fields (such as the inflaton $\varphi$ and/or the metric curvature).

We work with a spatially flat metric in conformal coordinates,
\begin{equation}
	ds^2=a^2 (\eta) \left( d\eta^2 - \delta_{ij} d x^i d x^j \right) \, ,
\end{equation}
with $\eta$ being the conformal time.  Adopting the longitudinal gauge $A_0 = 0$, $\partial^i A_i = 0$, for the vector potential, from (\ref{Lem}) one obtains the equation satisfied by the transverse field variable $A_i$:
\begin{equation} \label{eqAi}	
	A''_i  - \nabla^2 A_i + f' \epsilon_{ijk} \partial_j A_k = 0 \, , 
\end{equation}
where $\epsilon_{ijk}$ is the spatial Kronecker alternating tensor with $\epsilon_{123} = 1$, and the prime denotes the derivative with respect to the conformal time $\eta$.

In the spatial Fourier representation, and in the constant (in space and time) normalized helicity basis $\{ e^h_i (\bk) \}$, $h = \pm 1$, such that $\ri \bk \times \be^h = h k \be^h$, we have $A_i = \sum_h \cA_h e^h_i e^{\ri \bk \bx}$.  Then equation (\ref{eqAi}), for the helicity components $\cA_h$, implies
\begin{equation} \label{eqAh}	
	\cA''_h + \left( k^2 + h k f' \right) \cA_h = 0 \, . 
\end{equation}
The spectral densities of quantum fluctuations of magnetic and electric field are characterized, respectively, by the standard relations
\begin{align}
	\cP_B &= \frac{d \rho_B}{d \ln k} = \frac{k^4}{8 \pi^2 a^4} \sum_h \left| \cA_h (\eta, k) \right|^2 \, , \label{specB} \\
	\cP_E &= \frac{d \rho_E}{d \ln k} = \frac{k^4}{8 \pi^2 a^4} \sum_h \left| \frac{\cA'_h (\eta, k)}{k} \right|^2 \, , \label{specE}
\end{align}
in which the amplitude of the vector potential is normalized so that $\cA_h (\eta) \sim e^{- \ri k \eta}$ as $\eta \to - \infty$.  The factor in front of the sums in (\ref{specB}) and (\ref{specE}) is the spectral density of vacuum fluctuations in flat space-time in each mode at the physical wavenumber $k/a$.

Let the function $f$ evolve by power law during inflation \cite{Campanelli:2008kh}:
\begin{equation}\label{power}
	f (a) = f_0 \left( \frac{a}{a_\rf} \right)^p \, , \quad p \ne 0 \, ,
\end{equation}
where the index `f' denotes the end of inflation.  After the end of inflation, one can assume $f$ to quickly evolve to a constant value close to $f_0$.\footnote{Of course such an evolution of $f$ is supposed to be caused by its the dependence on the inflaton or other background fields.  For instance, in the case of inflation based on a massive scalar inflaton $\varphi$, one can assume $f \propto e^{- p \varphi^2 / M_\rP^2}$, where $M_\rP = (2 \pi G )^{-1/2}$ is a conveniently reduced Planck mass; see Eq.\@ (\ref{a-phi}) in Sec.~\ref{sec:inflaton}.}

During the quasi-exponential inflation (to which we restrict ourselves in this paper), the scale factor as a function of conformal time can be approximated as
\begin{equation}
	a (\eta) = - \frac{1}{H \eta} \, , \qquad \eta < 0 \, ,	
\end{equation}
where $H$ is the slowly varying Hubble parameter.  Treating $H$ as constant (i.e., considering the de~Sitter approximation), we can write equation (\ref{eqAh}) as
\begin{equation} \label{eqAh-c1}
	\cA''_h + \left[ k^2 + h k \frac{p f_0 }{| \eta_\rf |} \left( \frac{\eta_\rf}{\eta} \right)^{p + 1}  \right] \cA_h = 0 \, .
\end{equation}
A change of variable $\eta$ in this equation,
\begin{equation}\label{x}
	x = \frac{\eta}{\eta_\rf} \simeq \frac{a_\rf}{a} \, , 
\end{equation} 
puts it in the form
\begin{equation}\label{eqAh-c2}
	\cA''_h + \left( k^2 \eta_\rf^2 + \frac{h p f_0 | k \eta_\rf | }{x^{p + 1}} \right) \cA_h = 0 \, ,
\end{equation}
where the prime now denotes the derivative with respect to $x$, which evolves from large values to unity during inflation.  

For $x$ sufficiently large, so that the first term in the brackets of (\ref{eqAh-c2}) dominates, the normalized solution describing negative-frequency modes is given by
\begin{equation}\label{vac}
	\cA_h = e^{- \ri k \eta} = e^{- \ri k \eta_\rf x} \, , \qquad x \gg x_0 = \left| \frac{ p f_0}{k \eta_\rf } \right|^{1/(p + 1)} \, .	
\end{equation}

It is useful at this point to estimate the typical values of $|k \eta_\rf |$.  We have the following chain of relations:
\begin{equation}
	|k \eta_\rf | \simeq \frac{k}{a_\rf H_\rf } = \frac{k}{a_0 H_\rf } \cdot \frac{a_0}{a_\rf} \simeq \frac{k M_\rP}{a_0 g_r^{1/2} T_r^2} \cdot \frac{g_r^{1/3} T_r}{T_0} = \frac{k M_\rP}{a_0 g_r^{1/6} T_r T_0} \, .
\end{equation}
Here, $M_\rP = (2 \pi G )^{-1/2} = 4.8 \times 10^{18}\, \text{GeV}$ is a conveniently reduced Planck mass, $T_r$ is the reheating temperature, $T_0 = 2.34 \times 10^{-4}\, \text{eV}$ is the current temperature of the cosmic microwave background, and $g_r$ is the number of relativistic degrees of freedom in equilibrium after reheating.  In this estimate, we assumed reheating to proceed rapidly, so that the Hubble parameter $H_\rf$ at the end of inflation is roughly equal to that after reheating.  

If we are interested in characteristic wavenumbers of the order of $k / a_0 \simeq \text{Mpc}^{-1}$ (the corresponding spatial scale $\lambda = 2 \pi a_0 / k \sim 10\, \text{Mpc}$), then, for typical reheating temperatures $T_r \sim 10^{-4} M_\rP$ and $g_r \sim 10^3$, we obtain $|k \eta_\rf | \sim 10^{-22} \lll 1$.  Thus, for not too small values of $|p f_0|$, we have $x_0 \gg 1$.

\subsection*{The case of $p = 1$}

The case of $p = 1$ in (\ref{power}) was advocated in \cite{Campanelli:2008kh}.  In this case, equation (\ref{eqAh-c2}) has the exact solution in terms of the Hankel function of first kind,  
\begin{equation}\label{p1}
	\cA_h = e^{\frac{\ri \pi}{4} (\nu + 1 )} \sqrt{\frac{\pi | k \eta_\rf | x}{2}} H_{\nu/2}^{(1)} \left( |k \eta_\rf | x \right) = e^{\frac{\ri \pi}{4} (\nu + 1 )} \sqrt{- \frac{\pi  k \eta}{2}} H_{\nu/2}^{(1)} ( -  k \eta ) \, , 
\end{equation}
where
\begin{equation}\label{nub}
	\nu = \sqrt{1 + 4 b} \, , \qquad b = h f_0  k \eta_\rf \, ,
\end{equation}
which has correct asymptotics and normalization (\ref{vac}) as $x \to \infty$.  For the spectral density of electric field (\ref{specE}), we need to calculate its time derivative:
\begin{equation}\label{p1d}
	\frac{\cA'_h}{k} = e^{\frac{\ri \pi}{4} (\nu + 1 )}  \left[ \frac{\nu - 1}{2} \sqrt{- \frac{\pi}{2 k \eta}} H_{\nu/2}^{(1)} ( - k \eta ) - \sqrt{- \frac{\pi  k \eta}{2 }} H_{\nu/2 - 1}^{(1)} (- k \eta ) \right] \, . 
\end{equation}

The asymptotics of (\ref{p1}) and (\ref{p1d}) for $|k \eta| \ll 1$  and for real $\nu > 0$ are, respectively,
\begin{align}
	\cA_h &\approx e^{\frac{\ri \pi}{4} (\nu - 1 )} \frac{\Gamma (\nu / 2)}{\sqrt{\pi}} \left( - \frac{ k \eta}{2} \right)^{(1 - \nu)/2} \, , \\
	\frac{\cA'_h}{k} &\approx \frac{\nu - 1}{2} e^{\frac{\ri \pi}{4} (\nu - 1 )} \frac{\Gamma (\nu / 2)}{\sqrt{\pi}} \left( - \frac{ k \eta}{2} \right)^{- (1 + \nu ) / 2} \, .
\end{align}
From (\ref{specB}) and (\ref{specE}), we see that, in the long-wavelength regime, in each helicity,
\begin{equation}
	\frac{\cP_E}{\cP_B} = \left( \frac{\nu - 1 }{k \eta} \right)^2 \, .
\end{equation}
Since $|k \eta_\rf| \lesssim 10^{-22}$ for typical scales of interest, the energy density on such scales at the end of inflation would be extremely strongly dominated by electric field, causing the back-reaction problem, unless $\nu$ is very close to unity.  This last condition, $|1 - \nu| \ll 1$, will be satisfied if $|f_0|$ is not too large, specifically, if $|4 b| = |4 f_0 k \eta_\rf| \ll 1$.  In this case, we will have
\begin{align}
	\cP_B &= \frac{k^4}{8 \pi^2 a^4} \sum_h \left( - \frac{ k \eta}{2} \right)^{1 - \nu} \approx \frac{k^4}{8 \pi^2 a^4} \sum_h \left( - \frac{ k \eta}{2} \right)^{2b} \, , \\
	\frac{\cP_E}{\cP_B} &= \left( \frac{\nu - 1 }{k \eta} \right)^2 \approx \left( \frac{2 b}{k \eta} \right)^2 = \left( \frac{2 f_0 \eta_\rf}{\eta} \right)^2   \, .
\end{align}
Since $|b| = |f_0 k \eta_\rf| \ll 1$, on the spatial scales of interest at the end of inflation ($\eta = \eta_\rf$), we will have
\begin{align}
	\cP_B &\approx \frac{k^4}{8 \pi^2 a^4} \sum_h \left[ 1 + 2 h f_0 |k \eta_\rf| \ln | k \eta_\rf | + \left(2 f_0  k \eta_\rf \right)^2 \ln | k \eta_\rf | \right]   \nonumber \\ &= \frac{k^4}{4 \pi^2 a^4} \left[ 1 + \left(2 f_0 k \eta_\rf \right)^2 \ln | k \eta_\rf |  \right] \, , 
\end{align}
negligibly different from the spectrum of vacuum fluctuations (the first term in the last line of this expression).  

In the model considered in \cite{Campanelli:2008kh}, the quantity $f_0$ in (\ref{power}) was set to a scale-dependent expression $f_0 \propto k^{-p}$, making then $b$ and $\nu$ scale-independent constants.  This is inconsistent because $k$ is the wavenumber of the mode of electromagnetic field, while $f_0$ characterizes the evolution of spatially {\em homogeneous\/} coupling $f (\eta)$.  Thus, for $p = 1$, the quantity $f_0 \propto k^{-1}$ diverges as $ k \to 0$, producing the artificial infrared divergence in the electric spectral density if $|b| > 2$ (in the correct setting $f_0 = \text{const}$, from (\ref{nub}) we have $b \propto k$, and infrared divergence is absent). On the spatial scales of interest, the quantity $f_0$ was chosen in \cite{Campanelli:2008kh} as large as to make $0.1 \lesssim |b| \lesssim 2$.  In this case, one produces predominantly electric field, as we have just seen.

\subsection*{The case of $p =3$}

To study this case, consider equation (\ref{eqAh-c2}) at $x \ll x_0$, with the first term in the brackets neglected.  We have
\begin{equation}\label{eqAh-4}
	\cA''_h - \frac{b}{x^4} \cA_h = 0	\, , \qquad x \ll x_0 \, , \qquad b = 3 h f_0 k \eta_\rf  \, .
\end{equation}
Its general solution is
\begin{align}\label{sol4}
	\cA_h &= x \left( C_1 e^{\sqrt{b} / x} + C_2 e^{- \sqrt{b} / x}	\right) \, , \\
	\frac{1}{k} \frac{d \cA_h}{d \eta} &= \frac{1}{|k \eta_\rf|} \left[ C_1 e^{\sqrt{b} / x} \left( \frac{\sqrt{b}}{x} - 1 \right) - C_2 e^{- \sqrt{b} / x} \left( \frac{\sqrt{b}}{x} + 1 \right) \right] \, ,
\end{align}
where $C_1$ and $C_2$ are integration constants.  The gluing point, in view of (\ref{vac}), is located at
\begin{equation}
	x_0 = \left| \frac{3 f_0}{k \eta_\rf } \right|^{1/4} \gg 1 \, .
\end{equation}
Matching solutions (\ref{vac}) and (\ref{sol4}) at this point, we obtain
\begin{align}
	C_1 &= \frac{e^{- \sqrt{b}/x_0 + \ri \sqrt{|b|}/x_0}}{2 \sqrt{b}} \left(1 + \frac{\sqrt{b}}{x_0} - \ri \frac{\sqrt{|b|}}{x_0} \right)  \, , \\ 
	C_2 &= \frac{e^{\sqrt{b}/x_0 + \ri \sqrt{|b|}/x_0}}{2 \sqrt{b}} \left(- 1 + \frac{\sqrt{b}}{x_0} + \ri \frac{\sqrt{|b|}}{x_0} \right)  \, ,
\end{align}
where we have taken into account that $|k \eta_\rf | x_0 = \sqrt{|b|}/x_0$.

Again, assuming that $b$ is large and positive, for $x \ll x_0$, we have
\begin{equation}
	\cA_h \simeq C_1 x e^{\sqrt{b}/x} \, , \qquad \frac{1}{k} \frac{d \cA_h}{d \eta} \simeq \frac{1}{|k \eta_\rf|} C_1 e^{\sqrt{b}/x} \left( \frac{\sqrt{b}}{x} - 1 \right) \, ,
\end{equation}
so that, during inflation,
\begin{equation}
	\frac{\cP_E}{\cP_B} \simeq \left( \frac{\sqrt{b} - x }{k \eta_\rf x^2 } \right)^2 \simeq 10^{44} 
\end{equation}
for $k / a_0 \simeq \text{Mpc}^{-1}$.  Again we observe that this evolution predominantly produces electric field.

On the other hand, if $|b| \ll 1$, we have
\begin{align}
	C_1 &\approx \frac{1}{2 \sqrt{b}} \left[ 1 - \left( \frac{\sqrt{b}}{x_0} - \ri \frac{\sqrt{|b|}}{x_0} \right)^2 \right] \approx \frac{1}{2 \sqrt{b}} \, , \\
	C_2 &\approx - \frac{1}{2 \sqrt{b}} \left[ 1 - \left( \frac{\sqrt{b}}{x_0} + \ri \frac{\sqrt{|b|}}{x_0} \right)^2 \right] \approx - \frac{1}{2 \sqrt{b}} \, , 
\end{align}
and
\begin{align}
	\cA_h ( \eta_\rf ) &= C_1 e^{\sqrt{b}} + C_2 e^{-\sqrt{b}} \approx 1 \, , \\
	\frac{\cA'_h ( \eta_\rf )}{k} &\approx \frac{1}{k \eta_\rf}  \left( C_1 e^{\sqrt{b}} + C_2 e^{-\sqrt{b}} \right) \approx \frac{1}{k \eta_\rf} \, ,
\end{align}
without amplification of magnetic field but with a strong amplification of electric field.

The case of general $p > 0$ in (\ref{power}) can be analyzed in a similar way, but the above two examples are sufficient to show that it is problematic to amplify magnetic fields with evolution of the coupling in the form of positive power of the scale factor.

\section{A viable scenario}
\label{sec:viable}

The general mechanism of amplification of electromagnetic field in the present model is that, for sufficiently low values of $k$, the term $h k f'$ in the brackets of equation (\ref{eqAh}) dominates over $k^2$, and, for the helicity for which this term has negative sign, one  expects a regular growth of the corresponding mode.  In particular, this growth is exponential provided $f' = \text{const}$.\footnote{Evolution with constant $f'$ was under investigation in \cite{Hollenstein:2012mb} in the context of magnetogenesis in the hot universe, with negative conclusion as regards its efficiency essentially because of the high conductivity of cosmic plasma.} In this case, we have $f \propto \eta \propto a^{-1}$, so that now we deal with the evolution of the coupling in the form of {\em negative\/} power of the scale factor.

Basing on this observation, let us consider the scenario in which $f$ as a function of conformal time evolves as follows:
\begin{align}
	f (\eta) &= f_1 = \text{const} \, , \quad \eta \leq \eta_1  \, , \medskip \label{past}\\
	f' (\eta) &= \text{const} \, , \quad \eta_1 < \eta < \eta_2 \, , \medskip \\
	f (\eta) &= f_2 = \text{const} \, , \quad \eta \geq \eta_2 \, . \label{future}
\end{align}
In particular, the moment of time $\eta_2$ may mark the end of inflation.  The initially negative-frequency modes of the electromagnetic field evolve as
\begin{numcases}{\cA_h =}
	e^{- \ri k \eta}\, , &$\eta \leq \eta_1 \, ,$  \\ 
	a_k e^{- \ri k s_h \eta } + b_k e^{\ri k s_h \eta } \, , &$\eta_1 < \eta < \eta_2 \, ,$ \label{mid}  \\
	\alpha_k e^{- \ri k \eta } + \beta_k e^{\ri k \eta } \, , &$\eta \geq \eta_2 \, ,$  \label{fin}
\end{numcases}
where
\begin{equation}
	s_h = \sqrt{1 + h f' / k} \, .
\end{equation}
For $|f'|/k > 1$, the mode for one of the helicities in (\ref{mid}) will be hyperbolic, hence, exponentially amplified.  

By gluing the solutions at $\eta = \eta_1$ and $\eta = \eta_2$, we find the constants, respectively, in (\ref{mid}) and (\ref{fin}):
\begin{align}
	a_k &= \frac{s_h + 1}{2 s_h} e^{\ri k \eta_1 \left( s_h - 1 \right)} \, , \qquad  b_k = \frac{s_h - 1}{2 s_h} e^{- \ri k \eta_1 \left( s_h + 1 \right)} \, , \\
	\alpha_k &= e^{\ri k \Delta \eta} \left[ \cos \left( k s_h \Delta \eta \right) - \frac{\ri \left( s_h^2 + 1 \right)}{2 s_h} \sin \left( k s_h \Delta \eta \right)  \right] \, , \\
	\beta_k &= \frac{\ri \left( s_h^2 - 1 \right)}{2 s_h} e^{- \ri k \left( \eta_1 + \eta_2 \right)} \sin \left( k s_h \Delta \eta \right) \, ,
\end{align}
where $\Delta \eta = \eta_2 - \eta_1$.

After this evolution of the coupling, the mean occupation number of photons in a given mode is $n_k = \left| \beta_k \right|^2$.  Let us introduce the characteristic comoving wavenumber
\begin{equation}\label{km}
	k_{\rm m} = \frac12 |f'| \, ,
\end{equation}
and denote $\varepsilon_h = \text{sign} \left( h f'\right) = \text{sign} \left( h \Delta f \right)$.  Then, $s_h = \sqrt{1 + 2 \varepsilon_h k_{\rm m} / k}$.  Also take into account that $\Delta \eta = \Delta f / f'$, where $\Delta f = f_2 - f_1$. Then, depending on the values of $\varepsilon_h$ and $k$, we obtain the following expressions for the mean occupation numbers:
\begin{align}
	n_k	&= \frac{ \left( k_{\rm m} / k \right)^2}{ 1 + 2 \varepsilon_h k_{\rm m} / k } \sin^2 \left( \Delta f \frac{k}{2 k_{\rm m}} \sqrt{1 + 2 \varepsilon_h k_{\rm m} / k} \right) \, ,  &1 + \frac{2 \varepsilon_h k_{\rm m} }{k} > 0 \, , \label{harmon} \\
	n_k &= \frac{ \left( k_{\rm m} / k \right)^2}{2 k_{\rm m} / k - 1 } \sinh^2 \left( \Delta f \frac{k}{2 k_{\rm m}} \sqrt{2 k_{\rm m} / k - 1} \right) \, ,  &1 + \frac{2 \varepsilon_h k_{\rm m} }{k} < 0 \, , \label{hyper}
\end{align}
Note that the mean occupation numbers are functions only of $\Delta f$ and the ratio $k / k_{\rm m}$.  

\begin{figure}[htp]
	\centering
	\includegraphics[width=.7\textwidth]{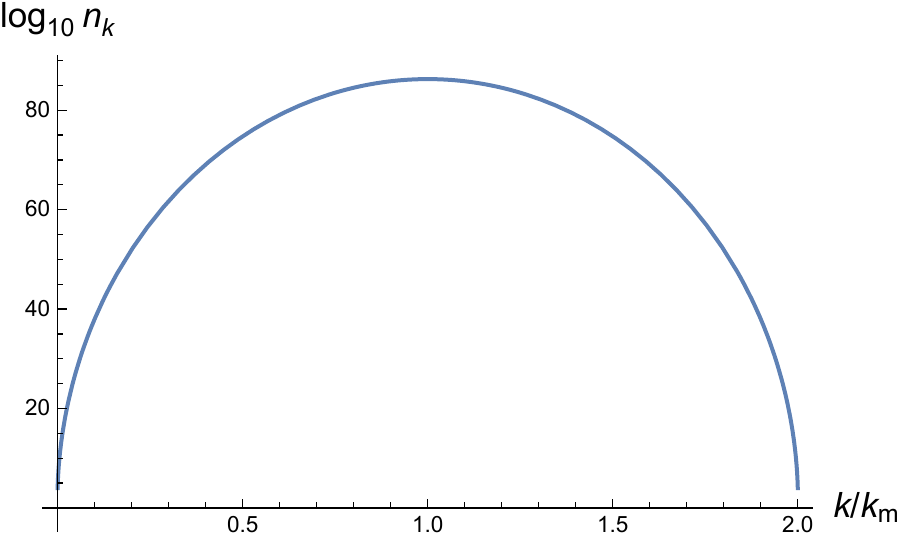}
	\caption{Spectrum (\ref{hyper}) on a logarithmic scale for $|\Delta f| = 200$. \label{fig:plot}}
\end{figure}

Spectrum (\ref{hyper}) for the helicity with $\varepsilon_h = -1$ is most interesting because it is exponentially peaked at the comoving  wavenumber $k = k_{\rm m}$ if $|\Delta f| \gg 1$.  At this peak, we have
\begin{equation}
	n_{\rm m}  = \sinh^2 \frac{\Delta f}{2} \, .
\end{equation}
At the same time, as $k \to 2 k_{\rm m}$ or $k \to 0$, we have $n_k \to ( \Delta f )^2 / 4$ for both helicities.  Spectrum (\ref{hyper}) is plotted on a logarithmic scale in Fig.~\ref{fig:plot} for $|\Delta f| = 200$.

The spectral densities (\ref{specB}) and (\ref{specE}) of fluctuations can be presented in the form
\begin{equation}\label{rBE}
	\cP_{B/E} = \frac{k^4}{4 \pi^2 a^4} \sum_h \left[ \frac12 + n_k \left( 1 \mp \zeta_k \right) \right] \, . 
\end{equation}
In the region $0 < k < 2 k_{\rm m}$, for the helicity with $\varepsilon_h = -1$, we have
\begin{align}
	\zeta_k = \left( 1 - \frac{k}{k_{\rm m}} \right) \cos \xi - \frac{k}{k_{\rm m}} \left( \frac{2 k_{\rm m}}{k} - 1 \right)^{1/2} \coth \left( \Delta f \frac{k}{2 k_{\rm m}} \sqrt{2 k_{\rm m} / k - 1} \right) \sin \xi \, ,
\end{align}
where $\xi = 2 k \left( \eta - \eta_2 \right)$.  At $\eta = \eta_2$, or if $k \left( \eta - \eta_2 \right) \ll 1$, neglecting the vacuum contribution, we then obtain
\begin{align}
	\cP_B &= \frac{k_{\rm m} k^3}{4 \pi^2 a^4} \frac{1}{2 k_{\rm m} / k - 1} \sinh^2 \left[ \Delta f \sqrt{ \frac{k}{2 k_{\rm m}} \left( 1 - \frac{k}{2 k_{\rm m}} \right)} \right] \, , \label{spectB} \\
	\cP_E &= \frac{k_{\rm m} k^3}{4 \pi^2 a^4} \sinh^2 \left[ \Delta f \sqrt{ \frac{k}{2 k_{\rm m}} \left( 1 - \frac{k}{2 k_{\rm m}} \right)} \right] \, . \label{spectE}
\end{align}
By making the Gaussian approximation to the spectra in this region, we can write
\begin{equation}
	\cP_B \simeq \cP_E \approx \frac{k_{\rm m}^4 e^{|\Delta f|}}{4 \pi^2 a^4} e^{- |\Delta f| \left( k - k_{\rm m} \right)^2/2 k_{\rm m}^2} \, , \qquad 0 < k < 2 k_{\rm m} \, . \label{Gauss}
\end{equation}
We observe that the spectral densities are peaked at the central value $k = k_{\rm m}$ with width $\Delta k = k_{\rm m} / \sqrt{|\Delta f|} \ll k_{\rm m}$.  The plots of these spectral densities, normalized to the vacuum value $k_{\rm m}^4 / 4 \pi^2 a^4$ of the spectral density on the scale $k_{\rm m}$, would be indistinguishable from that of Fig.~\ref{fig:plot}.  Thus, electric and magnetic fields are generated in this scenario with similar spectra in the spectral region of amplification.

Using approximation (\ref{Gauss}), one can estimate the total electromagnetic energy density: 
\begin{equation} \label{em}
	\rho_{\rm em} \simeq \frac{k_{\rm m}^4}{4 \pi^2 a^{4}} \sqrt{\frac{\pi}{2 | \Delta f|}} e^{| \Delta f |} \, .
\end{equation}
From this expression, we have an estimate for magnetic field:
\begin{equation}\label{main}
	B \simeq \frac{k_{\rm m}^2}{2 \pi a^2 |\Delta f|^{1/4}} e^{| \Delta f |/2} \, .
\end{equation}

Expressions (\ref{spectB})--(\ref{main}) contain two free parameters of the theory, $k_{\rm m} = | f' | / 2$ and $\Delta f$, which can easily be adjusted to produce magnetic fields of desirable strength with spectral density centered at the wavenumber $k_{\rm m}$.  (The bounds coming from the considerations of back-reaction on the inflationary dynamics will be derived in the next two sections.) For instance, in order to obtain $B_0 \simeq 10^{-15}\, \text{G}$ with spectrum peaked on the comoving scale $k_{\rm m}/a_0 \simeq \text{Mpc}^{-1}$ at the current epoch,\footnote{For simplicity, assuming the law $B \propto a^{-2}$ during all the time after generation, we neglect the possible chiral and turbulence effects \cite{Joyce:1997uy, Giovannini:1997eg, Boyarsky:2011uy, Saveliev:2013uva, Boyarsky:2015faa, Hirono:2015rla, Gorbar:2016qfh, Sidorenko:2016vom, Gorbar:2016klv, Pavlovic:2016gac, Brandenburg:2017rcb, Rogachevskii:2017uyc, Schober:2017cdw} that may modify the evolution of magnetic field.  The generated electric field will be washed-out by the plasma conductivity soon after reheating.} we require $|\Delta f| \simeq 200$.  Then $|\Delta f| = |f'| \Delta \eta = 2 k_{\rm m} \Delta \eta = 200$, hence $k_{\rm m} \Delta \eta \simeq 100$, i.e., evolution in $f (\eta)$ should start no later than about $\ln \left(k_{\rm m} \Delta \eta / k_{\rm m} \eta_\rf \right) = \ln 10^{24} \simeq  55$ $e$-foldings prior to the end of inflation.   The dependence of $|\Delta f|$ on the magnitude $B_0$ is quite weak (logarithmic); thus, for $B_0$ in the range $10^{- 30}$$-$$10^{- 7}$\,G on the same scale, one requires $|\Delta f| \simeq 131$$-$$238$.  

The ultraviolet oscillating `tail' in spectrum (\ref{harmon}) will be effectively exponentially cut without affecting the low-frequency part if we smooth-out the abrupt change in the derivative of $f (\eta)$ at $\eta = \eta_1$ and $\eta = \eta_2$, assuming that this process takes some time interval $\delta \eta \ll \Delta \eta \equiv \eta_2 - \eta_1$.  (For the case of kinetic coupling in (\ref{model}), this effect was nicely demonstrated in \cite{Campanelli:2016lxd}). Then the cut-off wavenumber $k_c \simeq 1 / \delta \eta$ and, according to (\ref{harmon}), the contribution from the `tail' to the electromagnetic energy density is $\rho_{\rm tail} \sim k_{\rm m}^2 k_c^2 / a^4$.  This contribution is negligible compared to (\ref{em}) if $k_c^2 \ll k_{\rm m}^2 | \Delta f|^{- 1/2} e^{| \Delta f|}$.  Since $k_{\rm m} = |f'|/2 = |\Delta f| / 2 \Delta \eta$, this condition translates into $\delta \eta / \Delta \eta \gg | \Delta f |^{-3/4} e^{- | \Delta f|/2} \sim 10^{- 45}$ for our typical value $|\Delta f| = 200$.

\section{Back-reaction on inflation}
\label{sec:inflation}

Note that the second term in (\ref{Lem}) has identically vanishing contribution to the stress-energy tensor (provided $f$ does not depend on the metric curvature, which is assumed to be the case).  Hence, all electromagnetic energy density stems from the first, canonical, term in (\ref{Lem}).

Suppose that the evolution of $f$ is completed $N$ $e$-foldings prior to the end of inflation (i.e., $a_\rf / a_2 = e^N$).  Let $B_\rf$ be the magnetic field by the end of inflation; then, since $B \propto a^{- 2}$ after generation, the magnetic field right after generation (at $\eta = \eta_2$) is $B_2 = B_\rf e^{2 N}$.  Note that, in our scenario, the  electromagnetic field reaches its maximum strength at the end of exponential amplification, i.e., by $\eta = \eta_2$.

Expressed through the extrapolated value $B_0$ of magnetic field at the current epoch, we have
\begin{equation}
	B_\rf = B_0 \left( \frac{a_0}{a_\rf} \right)^2 = B_0 \left( \frac{g_r}{2} \right)^{2/3} \left( \frac{T_r}{T_0} \right)^2 \, ,
\end{equation}
where, we remind the reader, $T_r$ is the reheating temperature, $T_0$ is the current temperature of the cosmic microwave background, and $g_r$ is the number of relativistic degrees of freedom in equilibrium after reheating.  The energy density of electromagnetic field at the end of the period of its generation is
\begin{equation}
	\rho_{\rm em}^{(2)} \simeq B_2^2 = B_\rf^2 e^{4 N} \simeq B_0^2 \left( \frac{g_r}{2} \right)^{4/3} \left( \frac{T_r}{T_0} \right)^4 e^{4 N} \, .
\end{equation}
This quantity should be smaller than the energy density of the universe during inflation, which, with the assumption of instantaneous reheating, is estimated as $\rho_{\rm inf} \simeq \pi^2 g_r T_r^4 / 30$.  This gives a back-reaction constraint
\begin{equation} \label{Bback}
	B_0 \ll g_r^{-1/6} e^{-2 N} T_0^2  \sim 10^{-6} g_r^{-1/6} e^{-2 N} \, \text{G} \, .
\end{equation}
Thus, it is possible to produce rather strong primordial magnetic fields with negligible back-reaction on inflation in this model if they are generated close to the end of inflation (i.e., if $N$ is not too large).  In the next section, we consider back-reaction on the inflaton dynamics in a typical inflationary model.

\section{Back-reaction on the inflaton}
\label{sec:inflaton}

To realize the suggested evolution of $f (\eta)$, in the approximation of almost constant $H$, we must assume
\begin{equation}
	f (\eta) = \text{const.} - \frac{f'}{a H} \, , \quad \eta_1 < \eta < \eta_2 \, .
\end{equation}
Then
\begin{equation}
	\Delta f =  \frac{f'}{a_1 H} - \frac{f'}{a_2 H} \approx \frac{f'}{a_1 H}
\end{equation}
if $a_1 \ll a_2$.  In this case, $f' = a_1 H \Delta f$, and
\begin{equation}\label{f-a}
	f (\eta) = \text{const.} - \frac{a_1 \Delta f}{a} \, , \quad \eta_1 < \eta < \eta_2 \, .
\end{equation}
In a particular model of inflation, we will have a relation $a = a (\varphi)$ during the slow-roll regime.  Equation (\ref{f-a}) then gives the necessary behavior of the coupling $f$ as a function of $\varphi$ in the specified time interval.  After that, one should ensure that $f$ becomes constant outside this time interval.  

Consider a simple example of a massive inflaton with potential $V (\varphi) = m^2 \varphi^2 / 2$.  In this case, 
\begin{equation}\label{a-phi}
	a (\varphi) \approx a_\rf e^{- \left( \varphi^2 - \varphi_\rf^2 \right) / M_\rP^2} \, ,	
\end{equation}
where, at the end of inflation, we have $\varphi_\rf^2 \approx M_\rP^2 / 3$.  Then an appropriate expression for $f (\varphi)$ is given by
\begin{equation}\label{f-inter}
	f (\varphi) = - \Delta f e^{\left( \varphi^2 - \varphi_1^2 \right) / M_\rP^2} \, , \quad \varphi_2 < \varphi < \varphi_1 \, ,
\end{equation}
where $\varphi_1 = \varphi (\eta_1)$ and $\varphi_2 = \varphi (\eta_2)$.  After inflation, the inflaton field $\varphi$ quickly relaxes to zero in the process of reheating, so that $f$ quickly becomes constant.  It is convenient then to associate $\eta_2$ with the end of inflation.  To make the function $f$ constant also for $\eta < \eta_1$, i.e., for $\varphi > \varphi_1$, we can modify (\ref{f-inter}) and suggest, for all values of $\varphi$,
\begin{equation} \label{f-2}
	f (\varphi) = - \Delta f \frac{e^{\left( \varphi^2 - \varphi_1^2 \right) / M_\rP^2}}{e^{\left( \varphi^2 - \varphi_1^2 \right) / M_\rP^2} + 1} = - \frac{\Delta f}{1 + e^{\left( \varphi_1^2 - \varphi^2 \right) / M_\rP^2}}\, .
\end{equation}
As the exponent $e^{\left( \varphi^2 - \varphi_1^2 \right) / M_\rP^2}$ evolves from large values at $\varphi > \varphi_1$ to small values at $\varphi < \varphi_1$, function (\ref{f-2}) evolves from a constant to the approximate behavior (\ref{f-inter}). To make the transition at $\varphi = \varphi_1$ even sharper, we can set
\begin{equation}
	f (\varphi) =- \frac{\Delta f}{\left[ 1 + e^{n \left( \varphi_1^2 - \varphi^2 \right) / M_\rP^2}\right]^{1/n}}
\end{equation}
with $n \gg 1$.  

The parameter $\varphi_1$ can be expressed through $k_{\rm m}$ and $\Delta f$.  Thus, from the estimates at the end of Sec.~\ref{sec:viable}, we know that setting the scale $k_{\rm m} / a_0 \sim \text{Mpc}^{-1}$ requires the evolution of $f$ to start about 55 $e$-foldings prior to the end of inflation.  Hence, in this case, we would require $\left( \varphi_1^2 - \varphi_\rf^2 \right) / M_\rP^2 \approx 55$, or $\varphi_1 \approx 7 M_\rP$.

The relative effect of back-reaction on the inflaton dynamics is estimated from Lagrangian (\ref{Lem}) and using (\ref{f-inter}) as (here, the primes denote the derivatives with respect to $\varphi$)
\begin{align}
	\left| \frac{\Delta V' (\varphi) }{V' (\varphi)} \right| = \frac{|f' (\varphi)|}{4 V' (\varphi)} \left| \left \langle F_{\mu\nu} \tilde F^{\mu\nu} \right\rangle \right| \simeq \frac{|f' (\varphi)|}{8 V' (\varphi)} \rho_{\rm em} \simeq \frac{|\Delta f|}{4 M_\rP^2 m^2} e^{\left( \varphi^2 - \varphi_1^2 \right)/ M_\rP^2} \rho_{\rm inf} \frac{\rho_{\rm em}}{\rho_{\rm inf}} \nonumber \\ \simeq \frac{|\Delta f| \varphi^2}{8 M_\rP^2} e^{\left( \varphi^2 - \varphi_1^2 \right)/ M_\rP^2} \frac{\rho_{\rm em}}{\rho_{\rm inf}} \lesssim \frac{|\Delta f| \varphi_\rf^2}{8 M_\rP^2} \frac{\rho_{\rm em}}{\rho_{\rm inf}} = \frac{|\Delta f| }{24} \frac{\rho_{\rm em}}{\rho_{\rm inf}} \, .
\end{align}
This is much smaller than unity if $\rho_{\rm em}/\rho_{\rm inf} \ll 24 / |\Delta f| \sim 10^{- 1}$ for our case $|\Delta f| = 200$, weakly strengthening our previous constraint ({\ref{Bback}) by a factor of one-third.  
	
\section{Implications for baryogenesis}
\label{sec:baryogenesis}

Staying in frames of the standard model of electroweak interactions, one can describe the post-inflationary evolution of the generated fields as follows.  During inflation, it is electromagnetic field which is generated because large quantum fluctuations of the Higgs field $\phi \sim H$ on super-Hubble spatial scales break the electroweak symmetry, leaving only the proton massless.\footnote{Unless interactions of the Higgs scalar with the metric curvature or inflaton generate large Higgs mass $m_\phi \gtrsim H$, preserving the electroweak symmetry during inflation.  In this case, one should consider generation of the weak-hypercharge gauge field.}  After reheating, the electroweak symmetry is restored, and only the weak-hypercharge part of the magnetic field survives on large spatial scales, making the field hypermagnetic.  This field evolves till the electroweak crossover at lower temperatures, during which it is gradually transformed to the usual magnetic field that survives until the present epoch \cite{Kajantie:1996qd, DOnofrio:2015gop}. 

One of the most interesting effects of the evolution of helical hypermagnetic fields is generation of baryon number in the early hot universe \cite{Giovannini:1997gp, Giovannini:1997eg}. This opens up an intriguing possibility of explaining the observed baryon asymmetry ($\eta_{\rm b} = n_{\rm b} / s \sim 10^{-10}$, where $n_{\rm b}$ is the baryon number density, and $s$ is the entropy density in the late-time universe) \cite{Bamba:2006km, Anber:2015yca, Fujita:2016igl, Kamada:2016eeb, Kamada:2016cnb, Jimenez:2017cdr}.  At the same time, this would rule out helical hypermagnetic fields that could overproduce this number.  According to the most elaborate recent calculations \cite{Kamada:2016cnb}, the resulting baryon asymmetry, when expressed through the present strength $B_0$ and correlation length $\lambda_0$ of (originally maximally helical) magnetic field, turns out to be
\begin{equation}\label{basym}
	\eta_{\rm b} \sim 10^{- (9 \text{--} 12)} \frac{\lambda_0}{\text{Mpc}} \left( \frac{B_0}{10^{-21}\, \text{G}} \right)^2 \, .
\end{equation} 
The uncertainty of about three orders of magnitude in this result is caused by the theoretical uncertainty in the dynamics of electroweak crossover in lattice simulations and analytical calculations.  Nevertheless, one can see that the present model of magnetogenesis can also support baryogenesis.  On the other hand, as follows from (\ref{basym}), to avoid overproduction of the baryon number, a model of magnetogenesis should respect a constraint on the current strength and correlation length of magnetic field, provided it was originally maximally helical and existed prior to the electroweak crossover:
\begin{equation}\label{bacon}
	B_0 \lesssim 10^{-21}\, \left( \frac{\text{Mpc}}{\lambda_0} \right)^{1/2} \, \text{G} \, .
\end{equation}
With $\lambda_0 \sim a_0/k_{\rm m}$, this constrains the possible values of $B_0$ and $k_{\rm m}$ in the present scenario.\footnote{Note that the physical quantity on the right-hand side of (\ref{basym}) evolves adiabatically, $\lambda B^2 \propto a^{-3}$, even with the effects of turbulence and inverse cascade taken into account \cite{Kamada:2016cnb}, so that its adiabatic extrapolation to the past is legitimate.}  

Constraint (\ref{bacon}) can probably be circumvented in frames of model (\ref{Lem}) by considering a non-monotonic evolution of $f$ which, after a period of time, reverses the sign of $f'$, thereby starting to amplify the mode with the opposite helicity.  The total magnetic helicity density $\rho_{\rm h}$ is given by
\begin{align}
	\rho_{\rm h} &= \frac{1}{a^3} \epsilon_{ijk} \left\langle A_i \partial_j A_k \right\rangle = \frac{1}{2 \pi^2 a^3} \int k^2 d k \sum_h h \left| \cA_h (\eta, k) \right|^2 \nonumber \\ &= \frac{1}{4 \pi^2 a^3} \int k d k \sum_h h \left[  | \beta_k |^2 + \Re \left( \alpha_k^{} \beta_k^* e^{- 2 \ri k \eta} \right)  \right] \, , \label{helicity}
\end{align}
where we have used expression (\ref{fin}) for the modes after inflationary amplification.  One can observe that reduction of the total helicity would require the mean occupation numbers $| \beta_k |^2$ at the end of evolution of $f$ to be (almost) the same for the two helicities in the spectral region of amplification,\footnote{This condition is realized, e.g., if the two subsequent periods of evolution have the same duration $\Delta \eta$ with constant derivatives $f'$ differing only by sign.} and the differences of the integrals of the second term in (\ref{helicity}) over the spectrum to be small.  Realization of such peculiar scenarios of non-monotonic evolution of $f$ does not look impossible but would probably require considerable fine-tuning.

Generating the baryon number, hypermagnetic fields also create baryon number {\em inhomogeneities\/} \cite{Giovannini:1997gp, Giovannini:1997eg} that can affect the cosmic microwave background and primordial nucleosynthesis.  This may introduce additional constraints on the magnetic field in the present model.  The problem of the relation between the power spectrum of baryon number inhomogeneities and the spectral density of helical hypermagnetic field is generic for the baryogenesis scenario under consideration and requires special investigation.

\section{Summary}
\label{sec:summary}

In this paper, we proposed a simple viable model of inflationary magnetogenesis based on the helical coupling in Lagrangian (\ref{Lem}).  In our case, the coupling $f$ evolves linearly with the conformal time $\eta$, interpolating between two constant values in the past and in the future (see ansatz (\ref{past})--(\ref{future})).  Contrary to the case of kinetic coupling, the absolute value of $f$ does not have any significance (since the second term in (\ref{Lem}) with constant $f$ is topological), so that the strong-coupling problem does not arise in this model.  The duration of the transition $\Delta \eta$ and the corresponding change $\Delta f$ are the two parameters of the model that can be adjusted to produce magnetic field of any strength in a narrow spectral band centered at any reasonable comoving wavenumber $k_{\rm m} = |f'| / 2 = |\Delta f| / 2 \Delta \eta$\,---\,see our formulas (\ref{spectB})--(\ref{main}) for the spectral density and strength of the field.  Constraint on the magnitude of the magnetic field comes from the considerations of back-reaction on the inflationary dynamics and, in the simple inflation based on a massive scalar field, allows for production of magnetic fields with extrapolated current values up to $B_0 \sim 10^{- 7}\, \text{G}$. The dependence (\ref{main}) of $|\Delta f|$ on $B_0$ is quite weak (logarithmic): for $B_0$ in the range $10^{- 30}$$-$$10^{- 7}$\,G with spectrum peaked on the comoving scale $k_{\rm m}/a_0 \simeq \text{Mpc}^{-1}$, one requires $|\Delta f| \simeq 131$$-$$238$.

In our scenario, the generated electromagnetic field is close to maximally helical, with the magnetic and electric fields having the same magnitude and spectral densities.  The electromagnetic spectrum is exponentially peaked around the comoving wavenumber $k_{\rm m} = |f'|/2$ with narrow width $\Delta k = k_{\rm m} / |\Delta f| \ll k_{\rm m}$ (see (\ref{Gauss})).  It is reasonable to think that evolution of the helical coupling in which $f'$ slowly varies in time will produce electromagnetic field with spectrum in broader range of wavenumbers, spanned by the values of $|f'|/2$.  

Soon after post-inflationary reheating, the electric field is washed-out by plasma conductivity, and the magnetic field, being maximally helical, may exhibit non-trivial subsequent evolution due to chiral and turbulence effects \cite{Joyce:1997uy, Giovannini:1997eg, Boyarsky:2011uy, Saveliev:2013uva, Boyarsky:2015faa, Hirono:2015rla, Gorbar:2016qfh, Sidorenko:2016vom, Gorbar:2016klv, Pavlovic:2016gac, Brandenburg:2017rcb, Rogachevskii:2017uyc, Schober:2017cdw} that may modify its power spectrum.  Primordial helical hypermagnetic fields may also be responsible for generating baryon asymmetry of the universe \cite{Bamba:2006km, Anber:2015yca, Fujita:2016igl, Kamada:2016eeb, Kamada:2016cnb, Jimenez:2017cdr}.  This imposes a post-inflationary constraint (\ref{bacon}) on the admissible values of $B_0$ and $k_{\rm m}$ in our simple scenario of monotonic evolution of $f$.  This constraint can probably be circumvented by assuming more complicated (non-monotonic) evolution of the coupling $f$ producing magnetic fields of sufficient strength but with conveniently limited helicity.  Other constraints on models of this type may arise from the considerations of the created baryon number inhomogeneities \cite{Giovannini:1997gp, Giovannini:1997eg} that can affect the cosmic microwave background and primordial nucleosynthesis.  This problem, specific to the discussed baryogenesis scenario, requires special investigation.  Another important issue that awaits for future analysis in the present scenario is the Schwinger effect of creation of charged particle-antiparticle pairs during magnetogenesis \cite{Sharma:2017eps, Kobayashi:2014zza, Sobol:2018djj}. 

\section*{Acknowledgments}

The author is especially grateful to Kohei Kamada for valuable communication on the theory of baryogenesis and related issues.  This work was supported by the National Academy of Sciences of Ukraine (project 0116U003191) and by the scientific program ``Astronomy and Space
Physics'' (project 19BF023-01) of the Taras Shevchenko National University of Kiev.

\end{document}